\documentclass[a4paper, amsfonts, amssymb, amsmath, reprint, showkeys, nofootinbib, twoside]{revtex4-1}
\usepackage{graphicx}  
\usepackage{dcolumn}   
\usepackage{bm}        
\usepackage{dutchcal}
\usepackage[toc,page]{appendix}
\usepackage{braket}
\usepackage{mathtools}

\begin{document}

\title{Constants of motion and quantum non-relativistic motion of a charged particle on a flat surface with transversal magnetic field}
\author{Gustavo V. L\'opez}
\email{gulopez@cencar.udg.mx}
\affiliation{Departamento de F\'isica, Universidad de Guadalajara, Guadalajara, M\'exico}
\author{Jorge A. Lizarraga}
\email{jorge\_lizarraga@icf.unam.mx\\}
\affiliation{Instituto de Ciencias F\'isicas, Universidad Nacional Aut\'onoma de M\'exico, Cuernavaca, M\'exico}
%
\date{\today}

\begin{abstract}
The motion of a charged particle moving on a flat surface is studied through the constants of motion associated to the system, given the magnetic gauge.
The usual Landau' solution  and the non separable solution for the Landau's gauge are found, and new  non separable solution for the symmetric gauge is given.
As a consequence of this, the quantization of the magnetic flux results from the invariance of the solutions under the unitary transformations which 
arise from the operators constants of motion.
\end{abstract}

\maketitle\noindent

\section{Introduction}
The problem of the quantum motion of a single charged particle $q$ of mass $m$ moving in a flat box under strong magnetic field, perpendicular to the flat motion and characterized by a vector potential ${\bf A}$, where ${\bf B}=\nabla\times{\bf A}$ 
 \cite{jack}, has received  attention during many years  \cite{Landau, lopez1,lau,cif} due to the finding of integer (IQHE) and fractional quantum Hall effect (FQHE) \cite{wk,kdp,tsg}, and to understand this new phenomenon \cite{amu,la2,ha,la3,tsg,la4,ja}.  
 The finding of a non separable solution of the problem has opened up a different way to focus the problem \cite{lop2}. We present here the solution of a single charged particle moving onlevel a flat surface from the perspective of the constant of motion associated to the system.  
We study the properties of the quantum system resulting from the classical Hamiltonian with a constant magnetic field based on the conserved operators to described the main characteristics of the system. 
\vskip2pc\noindent
We are interested in the Hamiltonian, which classically is written in CGS units as
\begin{equation}\label{H}
H=\frac{1}{2m}\left({\bf P}-\frac{q}{c}{\bf A}\right)^2,
\end{equation}
where $m$ is the mass of the particle, $q$ is the charge of the particle, $c$ is the speed of light, ${\bf P}=(p_{x},p_{y},p_{z})$ is the momentum of the particle and ${\bf A}=(A_{x},A_{y},A_{z})$ is the vector potential (also known as gauge) such that the magnetic field is given by  ${\bf B}=\nabla\times{\bf A}$. We set our magnetic field such that is parallel to the $z$ axis, ${\bf B}=B{\bf\hat k}$, where $B$ is constant. From now on we are only interested in the dynamics of the particle in the $x-y$ plane, therefore we set $p_{z}=0$ and $A_{z}=0$. 
Let us first use Landau's gauge defined as ${\bf A}_{L}=B(-y,0,0)$, thus, defining the cyclotron frequency\break $\omega_{c}=qB/mc$, the Hamiltonian (\ref{H}) is written as 
\begin{equation}\label{HL}
H_{L}=\frac{1}{2m}\left(p_{x}+m\omega_{c}y\right)^2+\frac{1}{2m}p_{y}^{2},
\end{equation}
and the Hamiltonian dynamical system is 
\begin{equation}\label{conservedLG1}
\frac{dx}{dt}=\frac{1}{m}(p_x+m\omega_cy)\quad\quad\frac{dp_{x}}{dt}=0,
\end{equation}
\begin{equation}\label{conservedLG2}
\frac{dy}{dt}=\frac{p_y}{m}\quad\quad \quad \frac{dp_y}{dt}=-\omega_c(p_x+m\omega_cy),
\end{equation}
which has the following conserved quantities (constants of motion)
\begin{equation}\label{cc1}
p_x,\quad\quad p_y+m\omega_cx,\quad\hbox{and}\quad H_L .
\end{equation}
This Hamiltonian system generates the equations of motion
\begin{equation}\label{newton_x}
m\ddot{ x}=m\omega_{c}\dot{ y},\quad\quad\hbox{and}\quad\quad
m\ddot{ y}=-m\omega_{c}\dot{ x},
\end{equation}
which are the same equations obtained by the Lorentz force, $m\ddot{\bf x}=\frac{q}{c}({\bf v}\times{\bf B})$, as expected. 
Their solutions  describe circular trajectories on the $x-y$ plane whose radius is $r=v/\omega_{c}$, were $v=\sqrt{v_{x}^{2}+v_{y}^{2}}$ is the speed of the charged particle.  
\vskip2pc\noindent
As it is well known, the Hamiltonian depends explicitly on the selection of the gauge. For instance, if one uses the symmetric gauge instead of Landau's gauge, \break ${\bf A}_{S}=\frac{B}{2}(-y,x,0)$, the Hamiltonian is written now as 
\begin{equation}\label{HS}
H_{S}=\frac{1}{2m}\left(p_{x}+\frac{m\omega_{c}}{2}y\right)^2+\frac{1}{2m}\left(p_{y}-\frac{m\omega_{c}}{2}x\right)^{2},
\end{equation}
and the Hamiltonian dynamical system is now given by
\begin{equation}\label{dxdtSG}
\frac{d x}{dt}=\frac{1}{m}\left(p_{x}+\frac{m\omega_{c}}{2}y\right),\quad \frac{dp_x}{dt}=+\frac{\omega_c}{2}\left(p_y-\frac{m\omega_c}{2}x\right)
\end{equation}
\begin{equation}\label{dydtSG}
\frac{d y}{dt}=\frac{1}{m}\left(p_{y}-\frac{m\omega_{c}}{2}x\right),\quad \frac{dp_y}{dt}=-\frac{\omega_c}{2}\left(p_x+\frac{m\omega_c}{2}y\right).
\end{equation}
resulting the following conserved quantities (constants of motion)
\begin{equation}\label{cc2}
p_{x}-\frac{m\omega_{c}}{2}y,\quad\quad p_{y}+\frac{m\omega_{c}}{2}x,\quad\quad\hbox{and}\quad H_S.
\end{equation}
It is not difficult to see that one gets the same equation of motion (\ref{newton_x}) for the charged particle motion evolution. 

\section{Quantum description of a charged particle under a constant magnetic field using the Landau's gauge}
To study the quantum dynamics, we asigne to the momentum variables the Hermitian operators $\hat{p}_k=-i\hbar\partial/\partial x_k$ with the conmutator relation $[{\hat x}_{a},{\hat p}_{b}]=i\hbar\delta_{a,b}$ and $[{\hat p}_{a},{\hat p}_{b}]=0$ (where ${\hat x}_{1}={\hat x}$, ${\hat x}_{2}={\hat y}$, ${\hat p}_{1}={\hat p}_{x}$ and ${\hat p}_{2}={\hat p}_{y}$), and solve the Schr\"odinger equation
\begin{equation}\label{sho}
i\hbar\frac{\partial\Psi}{\partial t}=\widehat{H}_L({\bf x},{\bf\hat p})\Psi,
\end{equation}
with the Hermitian Hamiltonian operator for the Landau's gauge (\ref{HL}) 
\begin{equation}\label{HL_Operator}
{\hat H}_{L}=\frac{1}{2m}\left({\hat p}_{x}+m\omega_{c}{\hat y}\right)^2+\frac{1}{2m}{\hat p}_{y}^{2}.
\end{equation}
In addition, the evolution of any  operator $\hat {f}$ can be seen in Heisenberg picture as
\begin{equation}\label{dfdt_Heisenberg}
i\hbar\frac{d{\hat f}}{dt}=[{\hat f},{\hat H}]+i\hbar\frac{\partial {\hat f}}{\partial t}.
\end{equation}
Using (\ref{dfdt_Heisenberg}) one can see that the operators associated to the constant of motion (\ref{cc1})
are also constant of motion for the quantum dynamics. Since they are time independent, one gets
\begin{equation}\label{conservedLG1_op}
[{\hat p}_{x},\widehat{H}_L]=0,
\end{equation}
\begin{equation}\label{conservedLG2_op}
[{\hat p}_{y}+m\omega_{c}{\hat x},\widehat{H}_L]=0,
\end{equation}
and, of course,
\begin{equation}\label{conservedLG3_op}
[{\widehat H}_L,\widehat{H}_L]=0.
\end{equation}
It is well known that (\ref{conservedLG1_op}) is used to solve  the Schr\"ondinger's equation and to have the so called Landau' solution of the problem. However,  the second conserved operator, (\ref{conservedLG2_op}) has been ignored so far \cite{Landau}, and we will show that both of them are needed to understand properly the system. It is not difficult to see that the analogous of 
(\ref{newton_x}) is gotten for the operators,
\begin{equation}\label{dxdtLG_op}
\frac{d^2 {\hat x}}{dt^2}=m\omega_c\frac{d{\hat y}}{dt},\quad\hbox{and}\quad
m\frac{d^{2} {\hat y}}{dt^{2}}=-m\omega_{c}\frac{d{\hat x}}{dt},
\end{equation}
Since the Hamiltonian (\ref{HL_Operator}) is time independent, we can propose on (\ref{sho}) a solution of the form 
\begin{equation}\label{time}
\Psi(x,y,t)=exp(-i E t/\hbar)\Phi(x,y)
\end{equation} 
which brings about the eigenvalue problem
\begin{equation}\label{eigen1}
\widehat{H}_L\Phi=E\Phi.
\end{equation}
We can also use the constants of motion (\ref{conservedLG1_op}) and (\ref{conservedLG2_op}) to build up a couple of eigenvalue equations 
\begin{equation}\label{eigenV_1}
{\hat p}_{x}\phi=-\lambda_1\phi,
\end{equation}
\begin{equation}\label{eigenV_2}
({\hat p}_{y}+m\omega_{c}{\hat x})\overline\phi=\lambda_2\overline\phi.
\end{equation}
Having the solutions
\begin{equation}\label{solution1}
\phi(x,y)=\psi_1(y)e^{-i\lambda_1 x/\hbar}
\end{equation}
and
\begin{equation}\label{solution2}
\overline\phi(x,y)=\psi_2(x)e^{i(\lambda_2-m\omega_cx)y/\hbar}.
\end{equation}
Substituting (\ref{solution1}) in (\ref{eigen1}), the resulting equation for $\psi_1$ is
\begin{equation}\label{ps1} 
\left[\frac{\hat{p}^2_{\xi}}{2m}+\frac{1}{2}m\omega_c^2\xi^2\right]\psi_1=E\psi_1,
\end{equation}
where $\xi=\lambda_1/m\omega_c+y$, obtaining the usual Landau' solution 
\begin{equation}\label{sol_1}
\Phi_n^{(1)}(x,y,\lambda_1)=e^{-i\lambda_1 x/\hbar}\psi_1^{(n)}\left(\sqrt{\frac{m\omega_c}{\hbar}}(\lambda_1/m\omega_c+y)\right),
\end{equation}
with the eigenvalues, so called Landau's levels,
\begin{equation}\label{ll}
E_n=\hbar\omega_c(n+1/2).
\end{equation}
The function $\psi_1^{(n)}$ is the normalized harmonic oscillator defined as 
\begin{equation}\label{harmonic_oscillator}
f_n(z)=\frac{1}{\sqrt{2^{n}n!}}\left(\frac{m\omega_{c}}{\pi\hbar}\right)^{1/4}e^{-\frac{1}{2}z^{2}}H_{n}(z),
\end{equation}
where $H_{n}(z)$ are the Hermite polynomials. \\\\
In addition, using (\ref{solution2}) in (\ref{eigen1}), the resulting equation for $\psi_2$ is
\begin{equation}\label{ps1} 
\left[\frac{\hat{p}^2_{\sigma}}{2m}+\frac{1}{2}m\omega_c^2\sigma^2\right]\psi_2=E\psi_2,
\end{equation}
where $\sigma=\lambda_2/m\omega_c-x$, and the solution is given by
\begin{eqnarray}\label{sol_2}
\Phi_n^{(2)}(x,y,\lambda_2)&=&e^{i(\lambda_2-m\omega_c x) y/\hbar}\times\nonumber\\
& &\psi_2^{(n)}\left(\sqrt{\frac{m\omega_c}{\hbar}}(\lambda_2/m\omega_c-x)\right),
\end{eqnarray}
where $\psi_2^{(n)}$ being the harmonic oscillator solution (\ref{harmonic_oscillator}), and with the same Landau's levels  (\ref{ll}) as eigenvalues. 
We want to point out that the wave function (\ref{sol_2}) is the  displaced solution found on reference \cite{lop2}. Therefore, 
the general solution of this system would be written as
\begin{equation}\label{G_sol_LG}
\Psi_{L}(x,y,t)=\sum_{n} e^{-\frac{i}{\hbar}E_{n}t}\left(a_n\Phi_{n}^{(1)}(x,y,\lambda_1)+b_n\Phi_{n}^{(2)}(x,y,\lambda_2)\right),
\end{equation}
where $a_n$ and $b_n$ are complex numbers. However, one need additional step to consider the degeneration 
of the system, and to do this, let us point out some symmetries of the system which are stablished by the constant of motion above. \\

As one could expected, these displaced functions (\ref{sol_1}) and (\ref{sol_2}) can be written as
\begin{equation}
\Phi_n^{(1)}(x,y,\lambda_1)=\widehat{U}_{\lambda_1}\Phi_n^{(1)}(x,y,0) 
\end{equation}
and
\begin{equation}
\Phi_n^{(2)}(x,y,\lambda_2)=\widehat{U}_{\lambda_2}\Phi_n^{(2)}(x,y,0),
\end{equation}
where the unitary operators $\widehat{U}_{\lambda_1}$ and $\widehat{U}_{\lambda_2}$ are defined as
\begin{equation}
\widehat{U}_{\lambda_1}=e^{-i(\lambda_1/\hbar m\omega_c)(\hat{p}_y+m\omega_c x)}
\end{equation}
and
\begin{equation}
\widehat{U}_{\lambda_2}=e^{-i(\lambda_2/\hbar m\omega_c)\hat{p}_y}.
\end{equation}
Because of the expressions (\ref{conservedLG1_op}) and (\ref{conservedLG2_op}), these operators commuted with the Hamiltonian and are elements of the group of symmetries 
of the Hamiltonian. In addition, it follows that
\begin{equation}
\widehat{H}_L(\hat{p}_y+m\omega_c x)\Phi_n^{(1)}=E_n(\hat{p}_y+m\omega_c x)\Phi_n^{(1)},
\end{equation}  
and
\begin{equation}
\widehat{H}_L\hat{p}_x\Phi_n^{(1)}=E_n\hat{p}_x\Phi_n^{(2)}
\end{equation}
that is, the application of the operators $\hat{\pi}_y=\hat{p}_y+m\omega_c x$ and $\hat{\pi}_x=\hat{p}_x$ on their respective functions, generates new eigenfunctions.  Defining these 
functions as
\begin{equation}\label{ffn}
f_n^j=\hat{\pi}_y^j\Phi_n^{(1)}\quad\hbox{and}\quad\tilde{f}_n^j=\hat{\pi}_x^j\Phi_n^{(2)},
\end{equation}
where the index "j" means the jth-application of the operator, and as we have just said
\begin{equation}
\widehat{H}_Lf_n^j=E_nf_n^j \quad\hbox{and}\quad\widehat{H}_L\tilde{f}_n^j=E_n\tilde{f}_n^j.
\end{equation}
This means that the most general solution is given by
\begin{equation}\label{gsol_1}
\Psi_{L}(x,y,t)=\sum_{n,j} e^{-\frac{i}{\hbar}E_{n}t}\left(a_{nj}f_n^j+b_{nj}\tilde{f}_n^j\right),
\end{equation}
where $a_{nj}$ and $b_{nj}$ are complex constants. If these constants are chosen as
\begin{equation}\label{guess}
a_{nj}=a_n\frac{1}{j!}\frac{\lambda_1^j}{(i\hbar)^j}\quad\hbox{and}\quad b_{nj}=b_n\frac{1}{j!}\frac{\lambda_2^j}{(i\hbar)^j},
\end{equation}
it happens that
\begin{equation}\label{guess_1}
\sum_ja_{nj}f_n^j=a_n\widehat{U}_{\lambda_1}\Phi_n^{(1)}(x,y,0)
\end{equation}
and
\begin{equation}\label{guess_2}
\sum_jb_{nj}\tilde{f}_n^j=b_n\widehat{U}_{\lambda_2}\Phi_n^{(2)}(x,y,0),
\end{equation}
which reproduces the solution (\ref{G_sol_LG}). 
So, one sees that the displaced solutions can be seen as a consequence of the degeneration of the system.
Of course, different selection of the constants (\ref{guess}) will result on different expression for (\ref{guess_1}) and (\ref{guess_2}). In addition,
defining the functions 
\begin{equation}
\Psi_1(x,y,t)=\sum_{n,j} e^{-\frac{i}{\hbar}E_{n}t}a_{nj}f_n^j
\end{equation}
and
\begin{equation}
\Psi_2(x,y,t)=\sum_{n,j} e^{-\frac{i}{\hbar}E_{n}t}b_{nj}\tilde{f}_n^j
\end{equation}
with the constants defined as (\ref{guess}), one can see the following effect for application of the unitary operators on these functions
\begin{equation}\label{symm1}
\widehat{U}_{\lambda_2}\Psi_1=e^{+i(\lambda_1\lambda_2/m\omega_c\hbar)}\Psi_1,
\end{equation} 
and
\begin{equation}\label{symm2}
\widehat{U}_{\lambda_1}\Psi_2=e^{-i(\lambda_1\lambda_2/m\omega_c\hbar)}\Psi_2,
\end{equation} 
which means that if $\Psi_1$ and $\Psi_2$ are invariant under these unitary transformations, the following relation must be satisfied
\begin{equation}\label{qflux_1}
\frac{\lambda_1\lambda_2}{m\omega_c\hbar}=2\pi k, \quad k\in  \mathbb{Z},
\end{equation} 
implying the quantization of the magnetic flux since \break $\omega_c=qB/mc$. Note that the units of $\lambda_1$ and $\lambda_2$ are the same 
units of the generalized linear momentun ($gr\cdot cm/sec$). Therefore, in this case  it is possible to propose $\lambda_1$ and $\lambda_2$ of the form
\begin{equation}\label{prop}
\lambda_1=m\omega_cl_1\quad\hbox{and}\quad \lambda_2=m\omega_cl_2,
\end{equation}
where $l_1$ and $l_2$ have units of length. In this way, (\ref{qflux_1}) would take the familiar form
\begin{equation}\label{qflux_2}
\frac{m\omega_c}{\hbar}l_1l_2=2\pi k, \quad k\in  \mathbb{Z}.
\end{equation}
Since the magnetic field is given (and therefore $\omega_c$), this expression suggests some type of quantization of the area defined by $l_1l_2$, see references \cite{lop3}.
%
\section{Quantum description of a charged particle under a constant magnetic field using the symmetric gauge}
For the symmetric gauge, ${\bf A}_{S}=B(-y,x,0)/2$, the Hamiltonian (\ref{HS})  is
\begin{equation}\label{HS_operator}
{\hat H}_{S}=\frac{1}{2m}\left({\hat p}_{x}+\frac{m\omega_{c}}{2}{\hat y}\right)^2+\frac{1}{2m}\left({\hat p}_{y}-\frac{m\omega_{c}}{2}{\hat x}\right)^{2}.
\end{equation}
The proposition(\ref{time}) reduced the study to solve an eigenvalue problem
\begin{equation}\label{eigHs}
\widehat{H}_S\Phi=E\Phi,
\end{equation}
and using (\ref{dfdt_Heisenberg}) it follows that
\begin{equation}\label{conservedSG1_op}
[{\hat p}_{x}-\frac{m\omega_{c}}{2}{\hat y},{\hat H}_{S}]=0,
\end{equation}
\begin{equation}\label{conservedSG2_op}
[\hat{p}_y+\frac{m\omega_{c}}{2}{\hat x}, {\hat H}_{S}]=0,
\end{equation}
and, of course,
\begin{equation}\label{conservedSG3_op}
[{\hat H}_{S}, {\hat H}_{S}]=0.
\end{equation}
These expressions are the operator equivalent expressions of the classical constants of motion (\ref{cc2}). Let us consider the eigenvalue problem for the operators (\ref{conservedSG1_op}) and 
(\ref{conservedSG2_op}),
\begin{equation}\label{eqS1}
({\hat p}_{x}-\frac{m\omega_{c}}{2}{\hat y})\varphi_1=\gamma_1\varphi_1
\end{equation}
and
\begin{equation}\label{eqS2}
(\hat{p}_y+\frac{m\omega_{c}}{2}{\hat x})\varphi_2=\gamma_2\varphi_2,
\end{equation}
 which have the solutions
 \begin{equation}\label{ssol_1}
 \varphi_1=\tilde{\psi}_1(y)e^{i(\gamma_1+m\omega_cy/2)x/\hbar}
 \end{equation}
 and
 \begin{equation}\label{ssol_2}
 \varphi_2=\tilde{\psi}_2(x)e^{i(\gamma_2-m\omega_cx/2)y/\hbar}
 \end{equation}
 Substituting (\ref{ssol_1}) and (\ref{ssol_2}) in (\ref{eigHs}) one gets the following equation for $\tilde{\psi}_1$ and $\tilde{\psi}_2$
 \begin{equation}\label{eqHS_1}
 \frac{\hat{p}_y^2}{2m}\tilde{\psi_1}+\frac{1}{2}m\omega_c^2\left(\frac{\gamma_1}{m\omega_c}+y\right)^2\tilde{\psi_1}=E\tilde{\psi_1}
 \end{equation}
 and
 \begin{equation}\label{eqHS_2}
 \frac{\hat{p}_x^2}{2m}\tilde{\psi_2}+\frac{1}{2}m\omega_c^2\left(\frac{\gamma_2}{m\omega_c}-x\right)^2\tilde{\psi_2}=E\tilde{\psi_2}
 \end{equation}
  which represent equations for the shifted quantum harmonic oscillator. Therefore, there are two independent solutions of the eigenvalue problem  (\ref{eigHs})

  \begin{eqnarray}\label{eigPhi1}
  \widetilde{\Phi}_n^{(1)}(x,y,\gamma_1)&=&e^{i(\gamma_1+m\omega_cy/2)x/\hbar}\times\nonumber\\
 & &\tilde{\psi}_1^{(n)}\left(\sqrt{\frac{m\omega_c}{\hbar}}(\gamma_1/m\omega_c+y)\right)
  \end{eqnarray}
 and
  \begin{eqnarray}\label{eigPhi2}
  \widetilde{\Phi}_n^{(2)}(x,y,\gamma_2)&=&e^{i(\gamma_2-m\omega_cx/2)y/\hbar}\times\nonumber\\
  & &\tilde{\psi}_2^{(n)}\left(\sqrt{\frac{m\omega_c}{\hbar}}(\gamma_2/m\omega_c-x)\right),
  \end{eqnarray}
 where $\tilde{\psi}_1^{(n)}$ and $\tilde{\psi}_2^{(n)}$ are the shifted functions (\ref{harmonic_oscillator}), and their eigenvalues are the same Landau's levels (\ref{ll}). We also have that these 
 shifted solution can be expressed through the unitary operator 
 \begin{equation}\label{HS_u1}
 \widehat{U}_{\gamma_1}=e^{-i(\gamma_1/\hbar m\omega_c)(\hat{p}_x-m\omega_c y/2)}
 \end{equation}
 and
 \begin{equation}\label{HS_u2}
 \widehat{U}_{\gamma_2}=e^{-i(\gamma_2/\hbar m\omega_c)(\hat{p}_y+m\omega_c x/2)}
 \end{equation}
 as
  \begin{equation}\label{u1_eig}
  \widetilde{\Phi}_n^{(1)}(x,y,\gamma_1)=\widehat{U}_{\gamma_2}\widetilde{\Phi}_n^{(1)}(x,y,0)
  \end{equation}
 and
  \begin{equation}\label{u2_eig}
  \widetilde{\Phi}_n^{(2)}(x,y,\gamma_2)=\widehat{U}_{\gamma_1}\widetilde{\Phi}_n^{(2)}(x,y,0)
  \end{equation}
 These unitary operators commute with the Hamiltonian and reflex the symmetries of the Hamiltonian. Thus, 
the general solution of this system, without the consideration of the degeneration, would be written as
\begin{equation}\label{G_sol_HS}
\Psi_{S}(x,y,t)=\sum_{n} e^{-\frac{i}{\hbar}E_{n}t}\left(\tilde{a}_n\widetilde{\Phi}_{n}^{(1)}(x,y,\gamma_1)+\tilde{b}_n\widetilde{\Phi}_{n}^{(2)}(x,y,\gamma_2)\right).
\end{equation}
To see the degeneration of the system, it is not difficult to see that due to commutation relations (\ref{conservedSG1_op}) and (\ref{conservedSG2_op}) one has
\begin{equation}
\widehat{H}_S\left({\hat p}_{x}-\frac{m\omega_{c}}{2} y\right)\widetilde{\Phi}_2^{(n)}=E_n\left({\hat p}_{x}-\frac{m\omega_{c}}{2} y\right)\widetilde{\Phi}_2^{(n)}
\end{equation}
and
\begin{equation}
\widehat{H}_S\left({\hat p}_{y}+\frac{m\omega_{c}}{2}x\right)\widetilde{\Phi}_1^{(n)}=E_n\left({\hat p}_{y}+\frac{m\omega_{c}}{2} x\right)\widetilde{\Phi}_1^{(n)}
\end{equation} 
 that is, the application of the operator $\tilde{\pi}_1=({\hat p}_{x}-\frac{m\omega_{c}}{2} y)$ on the eigenfunction $\widetilde{\Phi}_2^{(n)}$ and the application of the operator 
 $\tilde{\pi}_2=({\hat p}_{y}+\frac{m\omega_{c}}{2}x)$ on the eigenfunction $\widetilde{\Phi}_1^{(n)}$ generate new eigenfunctions. In this way, one can generate a numerable set 
 of eigenfunctions through the applications of these operators,
 \begin{equation}\label{ggn}
 g_n^j=\tilde{\pi}_2^j\widetilde{\Phi}_1^{(n)}, \quad\hbox{and}\quad \tilde{g}_n^j=\tilde{\pi}_1^j\widetilde{\Phi}_2^{(n)},
 \end{equation}
 getting now the general solution of the form
 \begin{equation}\label{gsol_2}
\Psi_{S}(x,y,t)=\sum_{n,j} e^{-\frac{i}{\hbar}E_{n}t}\left(c_{nj}g_n^j+d_{nj}\tilde{g}_n^j\right),
\end{equation}
 where $c_{nj}$ and $d_{nj}$ are complex number. Of course, one could select these complex coefficients as was done in previous case 
 and to make the addition with respect the degeneration index "j" to get solution (\ref{G_sol_HS}), and different selection will bring about different expressions. On the other hand, 
 one must note that due in general $\lambda_1\not=\gamma_1$ and $\lambda_2\not=\gamma_2$. Thus,
 the solution $\Psi_L$ ( with the Landau's gauge ${\bf A}= B(-y,0,0)$) and the solution $\Psi_S$ (with the symmetric gauge \break ${\bf A}=B(-y,x,0)/2$) can not be obtained each other through a gauge transformation.
 \\\\
Similarly to previous case, asking for the invariance of the functions, formed from (\ref{gsol_2}) considering only  the coefficients $c_{nj}$ or $d_{nj}$, and applying 
the respective unitary transformations (\ref{HS_u2}) and (\ref{HS_u1}), one would get the same phase, and their invariance of these function under the unitary transformation 
would lead us to a similar expression (\ref{qflux_1})
\begin{equation}\label{qflux_2}
\frac{\gamma_1\gamma_2}{m\omega_c\hbar}=2\pi k', \quad k'\in  \mathbb{Z},
\end{equation} 
and defining the constant $\gamma_2$ and $\gamma_1$ as
\begin{equation}
\gamma_1=m\omega_cl_1'\quad\hbox{and}\quad\gamma_2=m\omega_cl_2',
\end{equation}
having $l_1'$ and $l_2'$ units of length, an expression similar to (\ref{qflux_2}) is gotten
\begin{equation}\label{qflux_hs}
\frac{m\omega_c}{\hbar}l_1'l_2'=2\pi k', \quad k'\in  \mathbb{Z},
\end{equation}
where the quantized areas $l_1l_2$ and $l_1'l_2'$ must be the same since the quantization of the magnetic flux must be independent of the gauge chosen for the magnetic field.
\section{Conclusions and Comments}
We have studied the motion of a charged particle moving on a flat surface through the constants of motion associated to the system, and we have done this study 
with the Landau's gauge and the symmetric gauge for the constant magnetic field. The usual Landau' solution  and the non separable solution for the Landau's gauge were found, and a new  
non separable solution for the symmetric gauge was given. We also saw that, as a consequence of this, the quantization of the magnetic flux results from the invariance 
of the general solution under the unitary transformation which  are built  from the constant of motion operators, implying also the quantization some the area covered by the magnetic field.
We considere that the meaning of this resulting area quantization \cite{lop3} still needs to have further explanation and experimental verification. This invariance of the independent solution 
allows us to see the FQHF since the Hall resistivity is given by $\rho_{_H}=(\hbar/q^2)(m\omega_cl_1l_2/\hbar)$, but it does not allow us to see the IQHE, seen  reference \cite{lop3}.


\begin{thebibliography}{16}

\bibitem{jack}
Jackson, J.D. {\it Classical Electrodynamics}. John Wiley and Sons Inc., Hoboken, (1962). 
   
  \bibitem{Landau}
  Landau, L. D. and Lifshitz, E. M. {\it Quantum mechanics: non-relativistic theory (Vol. 3)}. Elsevier. (2013).
  
  \bibitem{lopez1}
  L\'opez, G. and Lizarraga, J.  {\it Charged Particle in a Flat Box with Static electromagnetic Field and Landau's leves},
  J. Mod. Phys. {\bf 11}, no.10, 1731-1742, (2020).
  
  \bibitem{lau}
Laughlin R.B.,  {\it Quantized motion of three two-dimensional electrons in a strong magnetic field},{Phys. Rev. B}, {\bf 27}, 3383-3389, (1983).

\bibitem{cif}
Cifja O., {\it Detailed solution of the problem of Landau states in a symmetric gauge}, {European Journal of Physics}, {\bf 41}, 035404, (2020).

\bibitem{wk} 
Wakabayashi, J.I. and Kawaji, S.  Journal of the Physical Society of Japan, 44,1839-1849, (1978).

\bibitem{kdp} 
Klitzing, K.V., Dorda, G. and Pepper, M. ,{\it New Method for High-Accuracy Determination of the Fine-Structure Constant Base on Quantized Hall Resistance} 
 Physical Review Letters , 45, (1980), 494-497.
 
\bibitem{tsg} 
 Tsui, D.C., Stormer, H.L. and Gossard, A.C. {\it Two-Dimensional Magnetotransport in the Extreme Quantum Limit}, Physical Review Letters , 48, (1982), 1559-1562. 

\bibitem{amu} 
Ando, T., Matsumoto, Y. and Uemura, Y.,  Journal of the Physical Society of Japan , 39, 279-288, (1975).

\bibitem{la2} 
Laughlin, R.B., Physical Review B , {\it Quantized Hall conductivity in two dimensions}, 
23, 5632(R), (1981).

\bibitem{ha} 
Halperin, B.I., Physical Review B , {\it Quantized Hall conductance, current-carrying edge states, and the existence states in two-dimensional disordered potential}, 
25, 2185, (1982).

\bibitem{la3} 
Laughlin, R.B., Uspekhi Fizicheskikh Nauk, 170, 292-303, (2000).

\bibitem{la4} 
Laughlin, R.B. (1983) Physical Review Letters , {\it Anomalous Quantum Hall Effect: An Incompressible Quantum Fluid with Fractional Charged Excitations}, 
50, 1395, (1983).

\bibitem{ja}
Jain, J.K., Physical Review Letters , {\it Composite-fermion approach for the fractional quantum {H}all effect}, 
63, 199-202, (1989).

\bibitem{lop2}
L\'opez, G. and Lizarraga J., J., Bravo J.P.O.,{\it Quantum Charged Particle in a Flat Box under Static Electromagnetic Field with Landau's gauge and Special Case 
with Symmetric Gauge}, 
 Mod. Phys. {\bf 12}, 1404-1414 (2021).

\bibitem{lop3}
L\'opez G. and Lizarraga J., J., {\it Single Charged Particle Motion in a Flat Surface with Static Electromagnetic Field and Quantum Hall Effect},  
Mod. Phys. {\bf 13}, 1324-1330, (2022).


\end{thebibliography}
\end{document}